\documentclass[fleqn,usenatbib]{mnras}

\usepackage{newtxtext,newtxmath}

\usepackage[T1]{fontenc}
\usepackage{comment}
\DeclareRobustCommand{\VAN}[3]{#2}
\let\VANthebibliography\thebibliography
\def\thebibliography{\DeclareRobustCommand{\VAN}[3]{##3}\VANthebibliography}

\usepackage{graphicx}	
\usepackage{amsmath}	

\title[Activity model in line profile time-series]{A Gaussian process model for stellar activity in 2-D line profile time-series}

\author[H. Yu et al.]{
Haochuan Yu$^{1}$\thanks{\href{mailto:haochuan.yu@physics.ox.ac.uk}{haochuan.yu@physics.ox.ac.uk}},
Suzanne Aigrain$^{1}$,
Baptiste Klein$^{1}$,
Michael Cretignier$^{1}$,
Florian Lienhard$^{2}$,
\newauthor
and Stephen J. Roberts$^{3}$
\\
$^{1}$Sub-department of Astrophysics, Department of Physics, University of Oxford, Oxford OX1 3RH, UK \\
$^{2}$Department of Physics, ETH Zurich, Wolfgang-Pauli-Strasse 2, CH-8093 Zurich, Switzerland \\
$^{3}$Machine Learning Research Group, Department of Engineering Science, University of Oxford, Oxford OX1 3PJ, UK
}

\date{Accepted XXX. Received YYY; in original form ZZZ}

\pubyear{2024}

\begin{document}
\label{firstpage}
\pagerange{\pageref{firstpage}--\pageref{lastpage}}
\maketitle

\begin{abstract}
Stellar active regions like spots and faculae can distort the shapes of spectral lines, inducing variations in the radial velocities that are often orders of magnitude larger than the signals from Earth-like planets. Efforts to mitigate these activity signals have hitherto focused on either the time or the velocity (wavelength) domains. We present a physics-driven Gaussian process (GP) framework to model activity signals directly in time series of line profiles or Cross-Correlation Functions (CCFs). Unlike existing methods which correct activity signals in line profile time series, our approach exploits the time correlation between velocity (wavelength) bins in the line profile variations, and is based on a simplified but physically motivated model for the origin of these variations. When tested on both synthetic and real data sets with signal-to-noise ratios down to $\sim$ 100, our method was able to separate the planetary signal from the activity signal, even when their periods were identical. We also conducted injection/recovery tests using two years of realistically sampled HARPS-N solar data, demonstrating the ability of the method to accurately recover a signal induced by a 1.5-Earth mass planet with a semi-amplitude of 0.3 m/s and a period of 33 days during high solar activity.
\end{abstract}



\section{Introduction}


Stellar variability is currently the limiting factor on the detection and characterisation of many exoplanets via the Radial Velocity (RV) method \citep[e.g.][]{2016Fischer,Crass2021}. This variability arises from the interplay between magnetic fields, convection and rotation. The most obvious component of this variability is associated with evolving magnetically active regions (spots and plage) on the rotating stellar surface, though (super-)~granulation in the quiet photosphere also plays a significant role \citep[e.g.][]{Meunier2015,Meunier2019,AlMoulla2023,OSullivan2024}. Over the past decade, considerable effort has been devoted to developing methods to model and mitigate these stellar RV signals in order to disentangle them from planetary signals \citep[see][for a review]{Zhao2022}. 

Most of the methods proposed so far can be grouped into three broad categories: a) time-domain methods, which model activity signals in RV time-series and, in some cases, a small number of additional activity indicators \citep[e.g.][]{2011A&A...528A...4B,2014MNRAS.443.2517H,V15,B22,D22}, b) velocity(wavelength)-domain methods, which typically work by separating pure Doppler shifts from line-shape variations in the Cross-Correlation Function (CCF) or full spectrum \citep[e.g.][]{CCameron2021,Zhao2022,deBeurs2022,Liang2024,Zhao2024,Klein2024}, c) line-by-line methods, which exploit the fact that different spectral lines respond to activity in a different manner, while Doppler shifts affect all lines in the same way \citep[e.g.][]{2018A&A...620A..47D,Cretignier(2020a),Cretignier2022,Shahaf2023,Lienhard2023}. Each of these approaches has shown promising results, but also suffers from limitations. Time-domain methods require good sampling of the stellar rotation, active region evolution, and planetary orbit timescales, and do not fully utilize the extensive information contained in the line-shape variations (and \emph{a fortiori} how these variations vary from line to line). Velocity-domain and Line-By-Line (LBL) methods, on the other hand, typically do not account for the time-dependence of the activity signals in general. In this paper, we introduce a hybrid method, which models both the time- and the velocity-dependence of activity signals explicitly, in time-series of line-profiles (or CCFs).

\begin{figure*}
	\centering
	\includegraphics[width=1.0\textwidth]{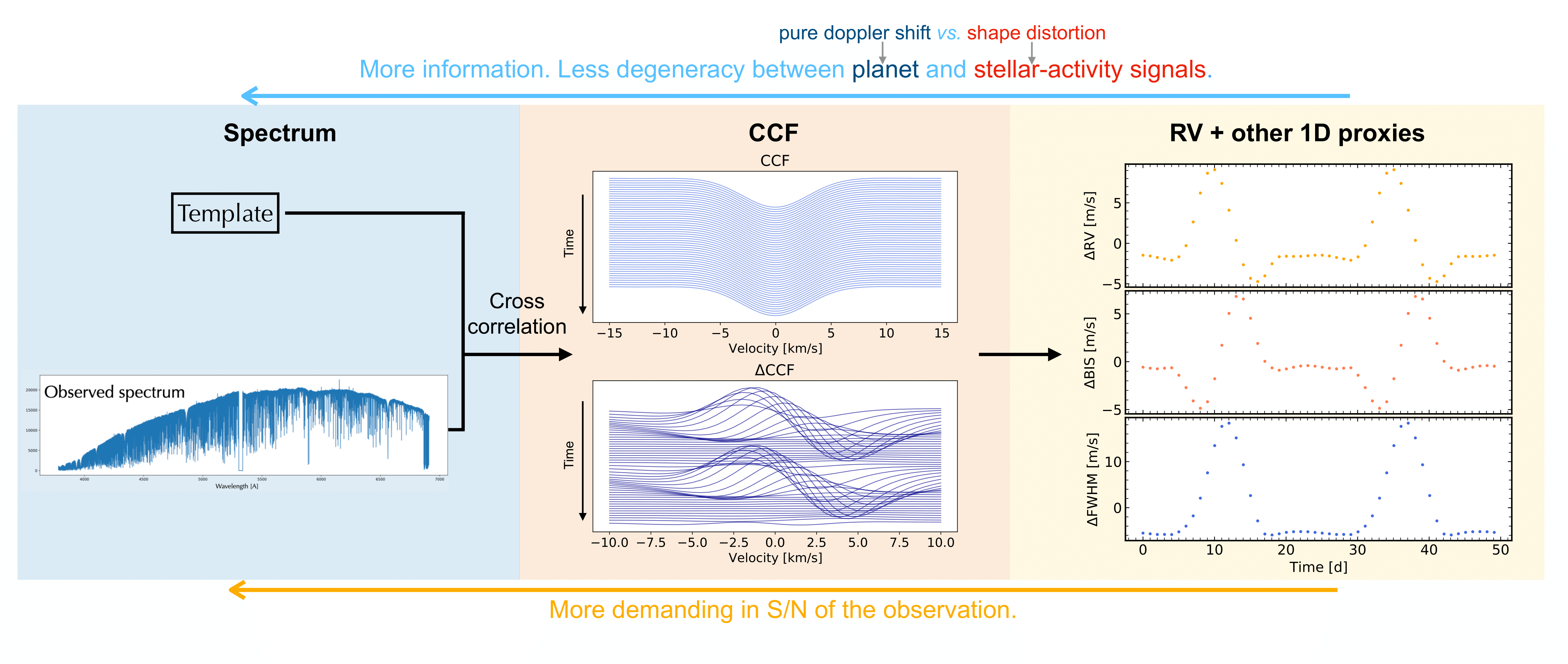}
    \caption{A typical workflow for RV extraction.}
    \label{fig:ccf-intro}
\end{figure*}

To place both existing methods and the new approach that we present here in context, it is helpful to recall the standard workflow for stellar RV measurements for ultra-stable spectrographs, which is illustrated in Figure \ref{fig:ccf-intro}. The workflow begins with a series of reduced, wavelength-calibrated, continuum-normalised high-dispersion spectra. In the most commonly used pipelines, such as the Data Reduction Software (DRS) used for the HARPS, HARPS-N, and ESPRESSO instruments \citep{Pepe2021}, each spectrum is cross-correlated with a template (usually, a digital binary mask constructed from a line list, \citealt{1996A&AS..119..373B,Pepe2002}), resulting in a CCF for each epoch. A simple function (usually a Gaussian) is then fit to the CCF, to derive the RV, together with the amplitude (contrast) and Full-Width at Half Maximum (FWHM) of the CCF. Additionally, a measurement of the CCF asymmetry, such as the Bisector Inverse Slope (BIS) is usually extracted at the same time. Furthermore, activity indicators such as the Mount Wilson $S$-index (or $\log R'_{\rm HK}$ computed from $S$-index) \citep{Noyes1984}, which quantify the amount of chromospheric emission in the cores of the Ca {\sc ii} H \& K lines, are frequently extracted from the spectra alongside the RVs and line-shape parameters. 

Alternatives to this workflow exist, particularly the LBL approach mentioned above, which bypasses the cross-correlation step and instead involves measuring separate RVs for thousands of individual spectral lines (the line lists used in LBL RVs are closely related to those used to construct the cross-correlation templates), or the template-free RV extraction method of \citet{2020MNRAS.492.3960R}, which works by pairwise comparison of the spectra from different epochs, within individual orders or segments thereof. The RVs for individual lines or segments are of course much less precise than the global RVs, but they can then be combined carefully to minimise sensitivity to instrumental effects and stellar signals.

This workflow, in all its various incarnations, can be seen as a dimensionality reduction procedure: each spectrum (consisting of $\sim$$10^5$ pixels) is reduced to a set of a few $10^3$ LBL or segment RVs, or to a CCF (consisting of a few tens to hundreds of velocity bins.), and then finally to a handful of numbers (one RV, plus a few line-shape and activity indicators). Stellar activity can be modelled and mitigated at any point in this process. Doing so at an earlier stage, when the dimensionality of the signal is larger, can be advantageous, as more information is retained, and the degeneracy between instrumental and stellar signals and pure Doppler shifts induced by companions is lower. However, this also requires that every spectrum has an excellent Signal-to-Noise Ratio (SNR), which is very expensive in terms of telescope time. Furthermore, these methods tend not to take the time-dependence of activity signals into account. On the other hand, the time-domain methods, which start from the lowest-dimension, end-result of the workflow, benefit from much more precise inputs (and hence have lower SNR requirements on the input spectra) and do use the time-domain information, but at the expense of discarding information about how activity signals affect the line shape and vary from line to line.

In this work, we therefore chose to start from an intermediate stage in the workflow, the line-profile or CCF stage. As the CCF combines information from thousands of spectral lines, its SNR is much higher than that of the spectra, so such a method should in principle be applicable to moderate SNR data. Most existing methods to model activity at the CCF level (\citealt{Jones2017}, \citealt{CCameron2021}, \citealt{John2022}, \citealt{Zhao2022}, \citealt{Klein2024}) work by separating the CCF variations which can be explained by pure Doppler shifts from those which correspond to changes in the profile shape. However, despite their diminished amplitude, notable quasi-periodic variations tied to stellar activity still persist in the shift-driven RVs.

These methods in general do not explicitly model the time-dependence of the line-profile variations. Doing this seems like a natural progression, as the geometry of the problem (evolving active regions on a rotating sphere) imparts fairly strong constraints on the joint time- and velocity-dependence of the activity signal. These constraints form the basis for the $FF'$ method of \citet{A12} and its extension to modelling RV and activity indicator time-series using GPs \citep{V15}, and here we extend that model to a 2--D time-series of line-profiles or CCFs. The hope is that the resulting model will provide a more complete description of the activity signal, while exploiting more of the information contained in the spectra, and being less sensitive to the time-sampling, than pure time-domain methods.

One subset of the existing approaches to model activity signals in time-series of line profiles does explicitly model their time-dependence: Doppler imaging \citep[e.g.,][]{Donati2014,Petit2015,Yu2017,Klein2021,Klein2022} seeks to reconstruct a map of the inhomogeneities on the stellar surface that reproduces the line-profile variations. Limitations of the approach include that it does not explicitly model the effects of convective blueshift suppression in active regions and does not account for the potential evolution of active regions over time. A recently developed \texttt{SpotCCF} framework \citep{DiMaio2024} use similar principles as Doppler imaging but restricts the number of spots in the reconstructed map of the stellar surface. This approach does allow flexibility for the spots to evolve, though do not specifically model the effects of convective blueshift suppression as well. The new method presented here addresses these limitations.

In this paper, we introduce the CCF-GP framework, a novel approach for modelling stellar activity signals within time-series of line-profiles or CCFs in Section \ref{sec:CCF-method}. We then present tests and applications of the framework on both synthetic and real data in Section \ref{sec:CCF-test}. We present our conclusions and suggest avenues for future work in Section \ref{sec:conc}. 

\section{A physics-based model for line profile variations} \label{sec:CCF-method}

We seek to model the perturbation to the mean profile of the absorption lines in a star's spectrum induced by evolving active regions on the rotating stellar surface. Unlike previous works, which modelled either the time- or the velocity(wavelength)-dependence of activity signals, but not both, we construct a 2-D model and apply it directly to the time-series of relative line profiles.

\subsection{Preliminaries}

Our model can be seen as an extension of the $F F^{\prime}$ framework introduced by \citet{A12}, which uses simple geometric arguments to elucidate the relationship between activity signals in photometry and in RV. We therefore start by reviewing the key elements of this framework.

\subsubsection{The $FF^{\prime}$ framework}

In the $F F^{\prime}$ framework, the quantity $F$ represents the relative variation in flux due to the presence of one or more active regions on the visible hemisphere of the star. $F$ is negative when there is a decrease in flux. Active regions can consist of dark spots, bright faculae, or a combination of the two. The rotation of the star causes the projected area of the active regions to change, and hence $F(t)$ to vary, where $t$ denotes time. Other factors also contribute to the time-dependence of $F(t)$, namely the intrinsic evolution (growth and decay) of the active regions, and the difference in limb-darkening properties between the active regions and the rest of the photosphere, but these are treated as second-order effects in the $FF^{\prime}$ framework.

The photometric contribution of any given active region to the total stellar RV variations scales as the product of $F(t)$, and 
$V_{\mathrm{loc}}(t)$, the line-of-sight velocity of the stellar surface at the location of the active region. 

In addition to this photometric effect, magnetically active regions also affect the total RV because a locally enhanced magnetic field density leads to a local reduction in the velocity of convective upflows. In the absence of active regions, these upflows result in a net convective blueshift, which is strongest near the disk centre and weakest near the limb. This blueshift is locally partially suppressed in the presence of active regions \citep[e.g.,][]{2008oasp.book.....G}. This "Inhibition of the Convective Blueshift" (ICB) effect is expected to scale as the product of the active region projected area, which varies like $F(t)$, and of the line-of-sight component of the up-flow velocity. 

The key insight of \citet{A12} is that, to first order, $V_{\mathrm{loc}}(t)$ behaves as the first derivative of $F(t)$, denoted by $F^{\prime}(t)$, while the line-of-sight component of the up-flow velocity behaves as $F(t)$. The total RV variation can therefore be approximated as the sum of two terms, one scaling as $F (t) F^{\prime}(t)$ and one scaling as $F^2(t)$. 

\subsubsection{The multi-GP framework}\label{sec:multi-GP}

\citet{A12} also noted that the time-dependence of $F(t)$ can be described by a GP with a quasi-periodic kernel. \citet{V15} then adapted the $FF^{\prime}$ formalism, to model simultaneously the signatures of active regions in RV time-series and in individual time-series of activity indicators extracted from the same spectra. These include both chromospheric indices such as the Mt Wilson $S$-index (or $\log R'_{\rm HK}$ computed from $S$-index), or line-shape indicators which trace changes in the width or asymmetry of the mean spectral line profile. Each time series is modelled as the linear combination of a term in $G(t) \equiv F^2(t)$ and a term in $G^{\prime}(t) \propto F(t) F^{\prime}(t)$, where $G(t)$ is a quasi-periodic GP, and the coefficients of proportionality are free parameters of the model. As this formalism involves a multi-output GP (where a single, latent GP variable is used to model multiple output time-series), we refer to it as the "multi-GP" approach. A number of extensions and public implementations of this framework have since been published \citep[e.g.][]{Jones2017,B22,D22,Nardiello2022,Hara2023}. 

\begin{figure*}
  \centering
  \includegraphics[width=1.0\textwidth]{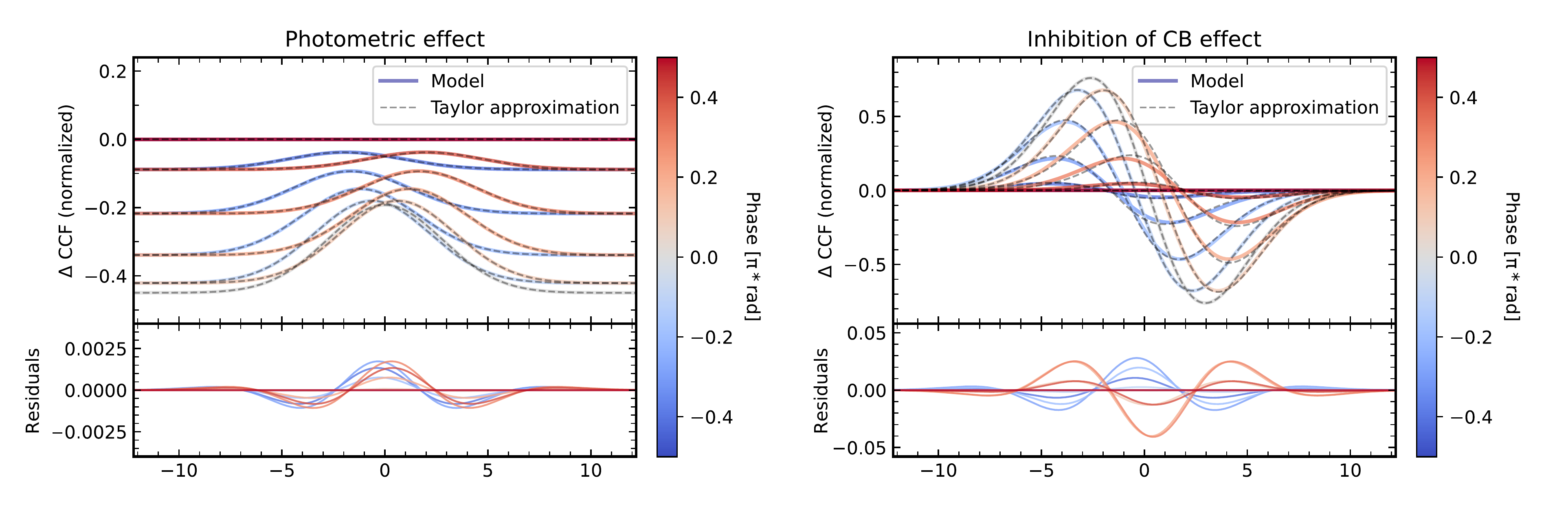}
    \caption{Perturbation of the line profile due to the photometric effect of a dark spot (left, $\Delta C_{\rm{phot}}$) and the ICB effect of a facula (right, $\Delta C_{\rm{icb}}$) for a single active region traversing the stellar disk along its equator. The coloured curves were evaluated according to Equations  \ref{eqa:ccf_p1} and \ref{eqa:ccf_icb1} respectively, at different rotational phases. The dashed grey lines show the corresponding Taylor approximations in Equations \ref{eqa:ccf_p2} and \ref{eqa:ccf_icb2}, truncated to $3^{\rm rd}$-order. The lower panels show the residuals between the line profiles evaluated from the original model and the Taylor approximations. In the left panel, the vertical offsets between $\Delta \rm{CCFs}$ from different rotation phases arise from the fact that we are estimating the line-profile using CCFs that are not continuum-normalised, and the continuum level therefore reflects the total flux variations induced by the active region. }
    \label{fig:ccf-1}
\end{figure*}

\subsection{Extension to line-profile variations}
\label{sec:extend}

We now consider, instead of a discrete set of time-series (RV and one or more activity indicators), a 2-D time-series of line profiles. More precisely, our starting point is a time-series of relative line-profiles, $\Delta C(v, t)$, where $v$ denotes velocity and $t$ denotes time, obtained by subtracting a reference line profile (representing the line profile in the absence of any active regions) $C_{\rm{ref}}(v, t)$ from the original time-series. In other words, we seek to model the perturbation to the line profile induced by (one or more) active regions. We do not specify at this stage how the line-profiles are extracted from the spectra, or how the reference line profile is constructed; this will be discussed in detail in Section~\ref{sec:implementation}. 

As in the $FF'$ framework, we treat the effect on the stellar line profile of an active region moving across the stellar disk as the sum of two distinct terms: the photometric effect and the ICB effect.

\subsubsection{The photometric effect}

To first order, an active region (e.g., a dark spot) alters the mean stellar line profile by suppressing the flux emerging from a small region of the stellar surface that rotates with the star. The resulting distortion of the line-profile, $\Delta C_{\rm{phot}}(v, t)$, can be modelled as:
\begin{equation} \label{eqa:ccf_p1}
\Delta C_{\rm{phot}}(v, t) \propto F(t) ~ C_{\rm{loc}}\left(v-V_{\mathrm{loc}}(t)\right),
\end{equation}
where $C_{\rm{loc}}(v)$ is the intrinsic line profile of the quiet photosphere (before rotational broadening). In Equation~\ref{eqa:ccf_p1}, it is evaluated at the location of the spot, i.e.\ Doppler-shifted by $V_{\mathrm{loc}}(t)$, the line-of-sight component of the stellar surface velocity at the location of the line profile. As in the $FF'$ framework, $F(t)$ is a time-dependent term that accounts for the spot's photometric contrast, fore-shortening, and limb-darkening. It can be thought of as the relative photometric perturbation caused by the spot, in the bandpass probed by the spectrograph. 

Following Equation~\ref{eqa:ccf_p1}, we introduce a Taylor-series expansion of $C_{\rm loc}$ about $V_{\rm loc}=0$:
\begin{equation} \label{eqa:ccf_p2}
\Delta C_{\rm{phot}}(v, t) \propto F(t) \, \sum_{k=0}^{\infty} \frac{V_{\mathrm{loc}}^k(t)}{k !} \, C_{\rm{loc}}^{(k)}(v),
\end{equation}
where $C_{\rm{loc}}^{(k)}(v)$ is the $k^{\rm th}$ derivative of $C_{\rm loc}$ with respect to $v$. The coloured curves in the left panel of Figure \ref{fig:ccf-1} show $\Delta C_{\rm{phot}}$ computed according to Equation \ref{eqa:ccf_p1} for a dark spot traversing the stellar disk along its equator. In this case, $F(t)$ is approximated as a pure sine function. The dashed grey lines show the Taylor approximation in Equation \ref{eqa:ccf_p2}, truncated to $3^{\rm rd}$ order.

As we have already seen, we can approximate $V_{\mathrm{loc}} (t)$ as $\gamma \, F^{\prime}(t)$, where $\gamma$ is a constant coefficient. Therefore, the line-profile perturbation due to the photometric effect can be represented as
\begin{equation} \label{eqa:ccf_p3}
\Delta C_{\rm{phot}}(v, t) =  \sum_{k=0}^{\infty} u_k(t) \, p_k(v),
\end{equation}
where the vectors $p_k(v) \equiv \mathrm{C}_{\mathrm{loc}}{ }^{(\mathrm{k})}(v)$ are given by the local, quiet line-profile and its derivatives, while the vectors $u$ are proportional to a combination of $F$ and its derivatives, similar to the original $FF^{\prime}$ framework:
\begin{equation}
\begin{aligned}
& u_0(t) \propto F(t), \\
& u_1(t) \propto \gamma F(t) F^{\prime}(t), \\
& u_2(t) \propto \frac{1}{2}  \gamma^2  F(t) F^{\prime 2}(t), \\
& u_3(t) \propto \frac{1}{6}  \gamma^3 F(t) F^{\prime 3}(t), ...
\end{aligned}
\end{equation} 

\subsubsection{Inhibition of convective blueshift effect}

As previously mentioned, the magnetic field associated with active regions suppresses convection locally, leading to a local redshift in the line profile. Note that this is a very simplified representation: in reality, the magnetic field also alters the line profiles. For now, we model the ICB effect as a straightforward shift in the local line profile at the location of the active region.

Given this premise, the perturbation of the line-profile due to the ICB effect can be expressed as
\begin{equation} \label{eqa:ccf_icb1}
\Delta C_{\rm{cb}}(v, t)  \propto F(t) \left[ C_{\rm{loc}}\left(v-V_{\mathrm{loc}}(t)-\delta V_{\rm{c}}\right) - C_{\rm{loc}} \left(v-V_{\rm loc}(t)\right) \right],
\end{equation}
where $\delta V_{\rm{c}}$ denotes the local redshift induced by the magnetic field. 
As in the case of the photometric effect, we then employ a Taylor-series expansion:
\begin{equation} \label{eqa:ccf_icb2}
\Delta C_{\rm{cb}}(v, t)  \propto F(t) \, \sum_{k=0}^{\infty} \frac{\left(V_{\rm{loc}}(t)-\delta V_{\rm{c}}\right)^k-V_{\rm {loc}}^k(t)}{k !} C_{\rm{loc}}^{(k)}(v).
\end{equation}
The right panel of Figure \ref{fig:ccf-1} shows the ICB perturbation to the line profile, $\Delta C_{\rm{cb}}$, caused by a facula crossing the stellar disk along its equator, as given by Equation \ref{eqa:ccf_icb1}. Again, here $F(t)$ is approximated as a pure sine function. The dashed grey lines show Taylor approximation of Equation \ref{eqa:ccf_icb2}, once again truncated to $3^{\rm rd}$ order.

We can thus represent the line-profile perturbation due to the ICB effect in the same form as in Equation \ref{eqa:ccf_p3}:
\begin{equation} \label{eqa:ccf_i3}
\Delta C_{\rm{cb}}(v, t) =  \sum_{k=0}^{\infty} u_k(t) \, p_k(v),
\end{equation}
but the vectors $u_k(t)$ are now given by 
\begin{equation}
\begin{aligned}
& u_0(t) = 0, \\
& u_1(t) \propto -\delta V_{\rm{c}} F(t), \\
& u_2(t) \propto -\delta V_{\rm{c}} \gamma F(t) F^{\prime}(t) + \frac{1}{2} \delta V_{\rm{c}}^2 F(t) , \\
& u_3(t) \propto -\frac{1}{2} \delta V_{\rm{c}} \gamma^2 F(t) F^{\prime 2}(t) + \frac{1}{2} \delta V_{\rm{c}}^2 \gamma F(t) F^{\prime}(t) - \frac{1}{6} \delta V_{\rm{c}}^3 F(t),\\
& ...
\end{aligned}
\end{equation}

\subsubsection{Full model}\label{sec:CCF-full}

For a single active region, we can combine the photometric and ICB components of the perturbation by adding them together. The full model takes the same form as Equations \ref{eqa:ccf_p3} and \ref{eqa:ccf_i3}:
\begin{equation} \label{eqa:ccf_pi}
\Delta C(v, t) =  \sum_{k=0}^{\infty} u_k(t) \, p_k(v),
\end{equation}
and, as before, the vectors $p_k(v) \equiv \mathrm{C}_{\mathrm{loc}}{ }^{(\mathrm{k})}(v)$ are given by the local, quiet line-profile and its derivatives,
but the vectors $u_k(t)$ are given by:
\begin{equation}\label{eq:ufull}
\begin{aligned}
& u_0(t) = \alpha  \, F(t), \\
& u_1(t) = \alpha  \, \gamma  F(t) F^{\prime}(t) - \beta \, \delta V_{\rm{c}}   F(t), \\
& u_2(t) = \frac{\alpha}{2}  \, \gamma^2   F(t) F^{\prime 2}(t) \\
& \quad \quad \ +\beta \, [-\delta V_{\rm{c}}  \gamma  F(t) F^{\prime}(t) + \frac{1}{2} \delta V_{\rm{c}}^2  F(t)] , \\
& u_3(t) = \frac{\alpha}{6}  \, \gamma^3  F(t) F^{\prime 3}(t)  \\
& \quad \quad \ +\beta \, [-\frac{1}{2} \delta V_{\rm{c}}  \gamma^2  F(t) F^{\prime 2}(t) + \frac{1}{2} \delta V_{\rm{c}}^2 \gamma  F(t) F^{\prime}(t) - \frac{1}{6} \delta V_{\rm{c}}^3   F(t)], \\
& ...
\end{aligned}
\end{equation}
where $\alpha$ and $\beta$ are constant coefficients. 

For the case where the line profiles are continuum-normalized, we expect $\Delta C_{\rm{norm}}(v, t)$ described by:
\begin{equation}\label{eq:cnorm}
\Delta C_{\rm{norm}}(v, t) = \frac{\Delta C(v, t) +C_{\rm{ref}}(v, t)}{1+ \alpha F(t)} - C_{\rm{ref}}(v, t),
\end{equation}
assuming the $C_{\rm{ref}}(v, t)$ is continuum-normalized already.

\subsubsection{Latent GP model for the $u_k(t)$} \label{sec:CCF-GP}

In the original multi-GP framework, which is tailored for 1D time-series data, each time-series is modelled as a sum of terms in $F^2$ and $F F'$. These are modelled by defining a latent variable $G \equiv F^2$, over a GP prior is placed. Because $dG/dt=2FF'$, the observed time series are linear combinations of the latent GP and its first derivative, and are thus GPs themselves. 

In our 2-D model for the line profile variations, the expressions for the vectors $u_k(t)$ in Equation~\ref{eq:ufull} involve combinations of $F$ and $F'$, $F'^2$, $F'^3$, \ldots\ These can no longer be expressed directly as linear combinations of a GP and its derivative(s). As a result, even if we assume that $F$, or some function of $F$ (such as $F^2$, $F'$ or $F'^2$) is a GP, the $u_k(t)$ are not GPs, and neither is the time-series of line-profile variations. 

We explored two possible solutions to this problem. The first consists in placing a GP prior on $F$ and jointly sampling $F$ and $F'$ from this prior, before computing $F'^2$, $F'^3$, \ldots. These are then used to compute the $u_k(t)$ before the resulting full model is compared to the observed line-profile time-series. This approach allows us to reproduce the precise form of Equation~\ref{eq:ufull}, but involves drawing a GP sample for every likelihood evaluation. The computational cost of such an approach proved to be prohibitive.

We therefore adopted a less accurate, but cheaper solution, which consists in treating each $u_k(t)$ as a linear combination of some GP $G(t)$ and its time-derivative $G^{\prime}(t)$, just as in the multi-GP framework described in Section~\ref{sec:multi-GP}. While this is clearly an approximation, our experience in modelling RV and activity indicator time-series with the multi-GP framework has shown that the combination of a GP and its first derivative is typically flexible enough to capture higher-order variations. 

Therefore, in our final model each $u_k(t)$ is approximated as $a_{k} G(t) + b_{k} G^{\prime}(t)$ and $p_k(v) \equiv \mathrm{C}_{\mathrm{loc}}{ }^{(\mathrm{k})}(v)$. We describe the procedure for deriving $p_k(t)$ from the observations later in Section~\ref{sec:implementation}. Consequently, the line profile residuals at a given velocity are also a linear combination of $G$ and $G^{\prime}$:
\begin{equation}
\Delta C(v,t) = \sum_{k=0}^r p_k(v) \, a_k \, G(t) + \sum_{k=0}^r p_k(v) \, b_k \, G^{\prime}(t),
\end{equation}
where $r$ is the number of terms used. We found that setting $r=3$ provides a sufficiently accurate approximation for most cases while also efficiently reducing computational costs.

Thus, the matrix $\boldsymbol{\Delta}\mathbf{C}$ is modelled as a latent GP model with $n$ outputs, where $n$ is the number of velocity bins, and the relationship between the output time-series is controlled by the $p_k(t)$, which corresponds to the emergent line profile of the quiet photosphere. Importantly, the total number of free parameters in the model is only, $2r+m$, where $m$ is the number of parameters of the GP covariance function used.

\subsubsection{Relationship to Principal Component Analysis}

The key insight in our model is that the line-profile perturbations can be modelled as a sum of separable functions of time and velocity. This is in some sense analogous to performing Principal Component Analysis (PCA) on the matrix of line-profile residuals. Specifically, the Singular Value Decomposition (SVD) of this matrix is given by
\begin{equation}
\boldsymbol{\Delta} \mathbf{C} = \mathbf{U} \, \mathbf{S} \, \mathbf{P},
\end{equation}
where $\mathbf{U}$ and $\mathbf{P}$ are rectangular matrices of left- and right-singular vectors (eigenvectors) which depend only on time and on velocity, respectively, while $\mathbf{S}$ is a diagonal matrix of singular values (square-root of eigenvalues). However, in SVD each set of singular vectors forms an orthonormal basis, whereas this isn't necessarily the case for the $u_k(t)$ and $p_k(v)$ vectors in our model.

\subsubsection{Limitations}

The physical model we have outlined in Section~\ref{sec:CCF-full} describes the perturbation to the profile of a single absorption line caused by a single active region crossing the stellar disk, and is in itself an approximation. Furthermore, there may be multiple active regions present at the same time at different locations on the stellar disk. The ratio of the $a$ and $b$ coefficients in Equation~\ref{eq:ufull} depends on the relative area and contrast spot and plage, which can differ between active regions, while the local redshift caused by plage ($\delta V_{\rm{c}}$ in Equation \ref{eqa:ccf_icb1}) is likely to depend on both the strength of the magnetic field in the plage and the viewing angle. Thus, treating the $\alpha$ and $\beta$ coefficients in Equation~\ref{eq:ufull} as independent of time is not strictly correct. 
Another aspect that we have overlooked in our physically-motivated model is that the observed line profile is not equal to the true line profile, but is convolved with the instrumental line profile. For instance, when we incorporate the distortion of the line profile $\Delta C$ into the reference profile $c_{\rm ref}$, both profiles have already been convolved with the instrumental profile. This can be slightly biased from the reality when the two un-convolved components are first summed, and the convolution is performed after.

To some extent, we expect these imperfections in our physical model to be absorbed by the flexible latent GP formulation we have adopted in Section~\ref{sec:CCF-GP}. Conversely, this means that the physical interpretation of the model becomes less tractable.

\subsection{Disentangling activity signals from planets}

\begin{figure*}
	\centering
	\includegraphics[width=1.0\textwidth]{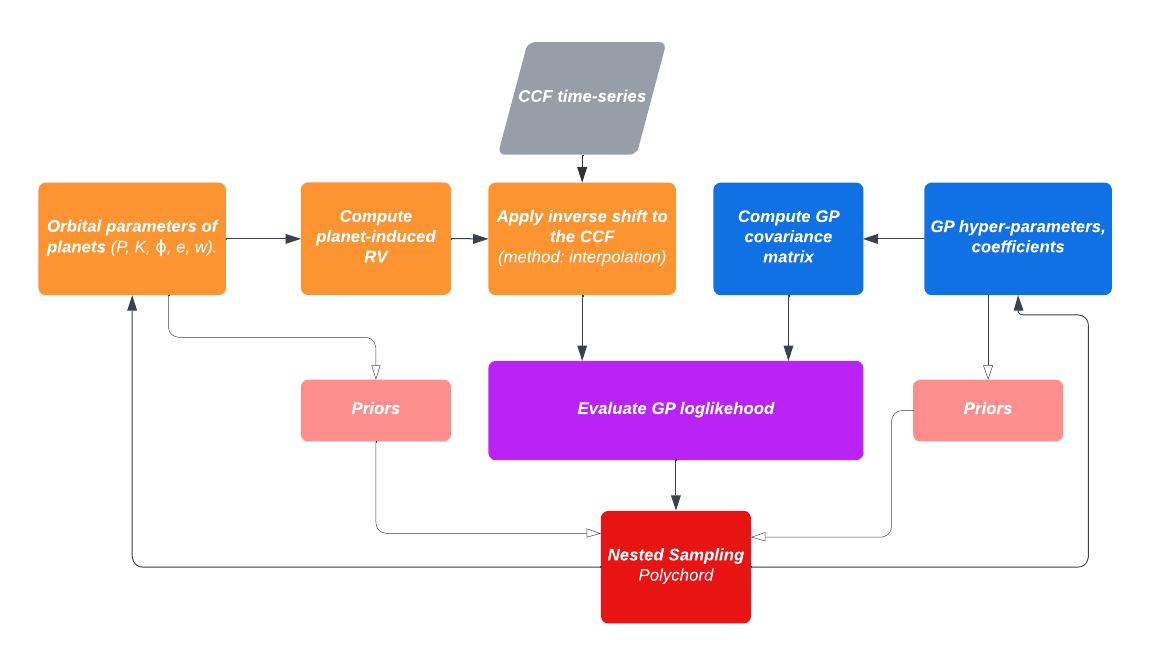}
    \caption{Flowchart illustrating the modelling workflow used in this work. The inputs are the observed time series of CCFs as well as the priors on all free parameters in the model. The output is the estimated posterior distribution of the free parameters.}
    \label{fig:ccf-workflow}
\end{figure*}

A key purpose of the model introduced above is to disentangle activity signals from the Doppler shifts caused by planetary companions, and thereby to improve the detectability and characterisation of the planets. To do this, we must explicitly account for the planet(s) when modelling the line-profile variations, as our model would otherwise have enough flexibility to absorb or modify some of the planet signals. The parameters of the planet(s) and the activity model must therefore be sampled jointly. In order to do this without escalating the computational cost, it is vital to keep the GP as the bottom layer of the workflow, to ensure the likelihood can still be computed analytically. 

\subsubsection{Workflow}

Figure \ref{fig:ccf-workflow} illustrates our overall workflow. At each step, we sample the parameters for the planet(s) and the activity model simultaneously. The $5p+1$ planet parameters (where $p$ is the number of planets, and the remaining parameter defines the systemic RV) are used to evaluate the planet-induced RV signal, and the line profiles are then shifted inversely by the corresponding amount.

To apply a shift to the line profiles, we initially fit a non-parametric function to each profile using least-squares cubic interpolation, as implemented in \texttt{SciPy} \citep{2020SciPy-NMeth}\footnote{\url{https://scipy.org/}}. Subsequently, we apply an inverse shift to the velocity grid of each line profile to account for the RV signal induced by the planet. This shifted velocity grid is then input into the fitted non-parametric function to compute the shifted line profile for each time stamp.


In parallel, given the $2r+m$ activity parameters, a GP solver is used to evaluate the covariance matrix $\mathbf{K}$ of the model and compute a marginal likelihood according to:

\begin{equation}
\log \mathcal{L}(\boldsymbol{\theta}, \boldsymbol{\phi})=-\frac{N_{\rm {obs}}}{2} \log (2 \pi)-\frac{1}{2} \log |\boldsymbol{K}|-\frac{1}{2} \boldsymbol{r}^{\mathrm{T}} \boldsymbol{K}^{-1} \boldsymbol{r}
\end{equation}
where $\boldsymbol{\theta}$ and $\boldsymbol{\phi}$ are the parameters of the planet and activity models, $\boldsymbol{r}=\boldsymbol{y}-\boldsymbol{\mu}$ contains the residuals of the observed line profiles $\mathbf{y}$ subtracted by the reference (mean) line profile $\boldsymbol{\mu}$, `flattened' into a one-dimensional vector. $N_{\rm{obs}}$ is the number of observations. As we are modelling relative line profiles, which have been shifted to account for the planet signal, the mean function of the GP is zero everywhere.

\subsubsection{Practical implementation of GP}

We use the fast GP solver \texttt{S+LEAF 2}\footnote{\url{https://gitlab.unige.ch/Jean-Baptiste.Delisle/spleaf}} \citep{D22}
with a Mat\'ern 3/2 exponential periodic (MEP) kernel. This serves as a fast approximation of the quasi-periodic kernel frequently used to describe rotational modulation active regions in stellar light curves and RV time-series, allowing the likelihood evaluation to scale linearly with the number of observations and velocity bins. We also include a constant term on the diagonal of the covariance matrix to absorb any imperfections in our model, which results in $m=4$ parameters for the GP covariance function. 

As we are sampling over a relatively large number of parameters, and expect the likelihood hyper-surface to be multi-modal in some cases, we sample the joint posterior over the parameters using the \texttt{PolyChord}\footnote{\url{https://github.com/PolyChord/PolyChordLite}} nested sampler \citep{2015MNRAS.450L..61H,2015MNRAS.453.4384H}. Nested sampling is generally considered more adept at handling multi-modal problems, making it a more appropriate choice for this application than more standard Markov Chain Monte Carlo (MCMC) samplers, while the polychord implementation is known to perform especially well for large numbers of parameters.

\subsection{Model inputs} \label{sec:implementation}

The starting point for our method is a matrix $\boldsymbol{\Delta} \mathbf{C}$ containing a time-series of relative line-profiles:
\begin{equation}
    \Delta C_{ij} = C_{ij} - c_{\rm ref,i} = C\left(v_i,t_j\right) - c_{\rm ref}  (v_i),
\end{equation}
obtained by subtracting a reference line profile $\mathbf{c}_{\rm ref}$ (representing the line profile in the absence of any active regions) from each row of the matrix of line profiles $\mathbf{C}$.  

\subsubsection{Extracting the line profiles}

Except where stated otherwise, we equate each row of $\mathbf{C}$ to the CCF produced by the standard "DRS" (Data Reduction Software, used for
HARPS, HARPS-N and ESPRESSO data (e.g., \citealt{Lovis2007}) for that spectrum. These CCFs are obtained by cross-correlating the spectra with a binary mask, which is zero everywhere except at the centre of individual spectral lines (\citealt{1996A&AS..119..373B}, resulting in "inverted" CCFs, which resemble an absorption line profile, e.g.\ the top-middle panel in Figure~\ref{fig:ccf-intro}). Note that if the CCFs are un-normalised, the continuum level (away from the centre of the line profile) varies with time. 

In principle, it may be preferable to extract line profiles from the spectra using Least Squares Deconvolution (LSD, e.g., \citealt{Donati1997,2010A&A...524A...5K}). However, we chose to use CCFs in this work because they are computed automatically by the DRS and thus readily available. In practice, we have found that, provided the line lists used in the creation of the CCF template and in the LSD process are similar, the resulting line profiles are also very similar. 

\subsubsection{Reference profile}
Ideally, the reference profile $\mathbf{c}_{\rm ref}$ should be equal to the line profile in the absence of any active regions, so that the residuals $\boldsymbol{\Delta} \mathbf{C}$ represent the perturbations induced by activity. In practice, we do not normally have spectra which are guaranteed to be activity-free. Therefore, the reference line profile is obtained by averaging the line profiles $\mathbf{C}$ over the time axis, optionally selecting a subset with a lower activity level (as measured, for example, by the calcium $S$ or $\log R'_{\rm HK}$ indices).

\subsubsection{Estimating the $p_k(v)$}
We approximate the local, quiet line profile $\mathrm{C}_{\mathrm{loc}}{ }^{(k)}(v)$ as the deconvolution of the reference profile $\mathbf{c}_{\rm ref}$ with a rotational broadening kernel estimated from the $v\sin i$ of the star. The effect of the rotational broadening becomes negligible for slow rotators, when the $v \sin i$ is significantly smaller than the width of the instrumental profile.

\section{Tests and applications} \label{sec:CCF-test}
In this section, we analysed both synthetic and real datasets. Our objectives are fourfold: a) to validate the framework by assessing its ability to model signatures from stellar spots and faculae; b) to determine if the framework can more effectively disentangle planetary and stellar signals than existing 1-D time series methods; c) to evaluate its efficiency in detecting planets with low-amplitude signals; and d) to demonstrate its capability to operate effectively at a Signal-to-Noise Ratio (SNR) typical of most real-world observations, approximately 100.

\subsection{Simulated data from SOAP 2.0} \label{sec:CCF-test-soap}
Before implementing the suggested framework on real data, we tested it using simulated data from version 2 of the \texttt{SOAP} (Spot Oscillation and Planet) code \citep{Dumusque2014,Boisse2012}. \texttt{SOAP2}~\footnote{\url{https://www.astro.up.pt/resources/soap2/}} enables users to generate time-series of CCFs distorted by active regions, namely spots and faculae, on the stellar surface. This is achieved by specifying the latitude, longitude, and size of each region, alongside the star's basic properties such as its rotation period. The tests involved two configurations -- cases <i> and <ii>, detailed in Table \ref{tab:soap2}, including the simulated rotation period of the star, the numbers and locations of the spots and faculae. For both cases, the sampling rate is one point per day, and the 150-day duration corresponds to six times the rotation period of the star.

\begin{table}
    \centering
    \caption{Key input parameters for simulating time-series of CCFs with \texttt{SOAP2}.}
    \label{tab:soap2}
    \begin{tabular}{cccccccc}
       \hline
       Case  & $P_{\rm{rot}}$ & N$_{\rm{spots}}$ & Long & Lat &  N$_{\rm{faculae}}$ & Long & Lat \\
       -- & [d] & -- & [deg] & [deg]  & -- & [deg] & [deg]\\
       \hline
       <i> & 25.0 & 1 & 180 & 30 & 1 & 60 & 60 \\
       <ii> & 25.0 & 2 & 180,-150 & 30,-15 & 2 & 60,-40 & 60,70 \\
       \hline
    \end{tabular}
\end{table}

\begin{figure*}
	\centering
	\includegraphics[width=\textwidth]{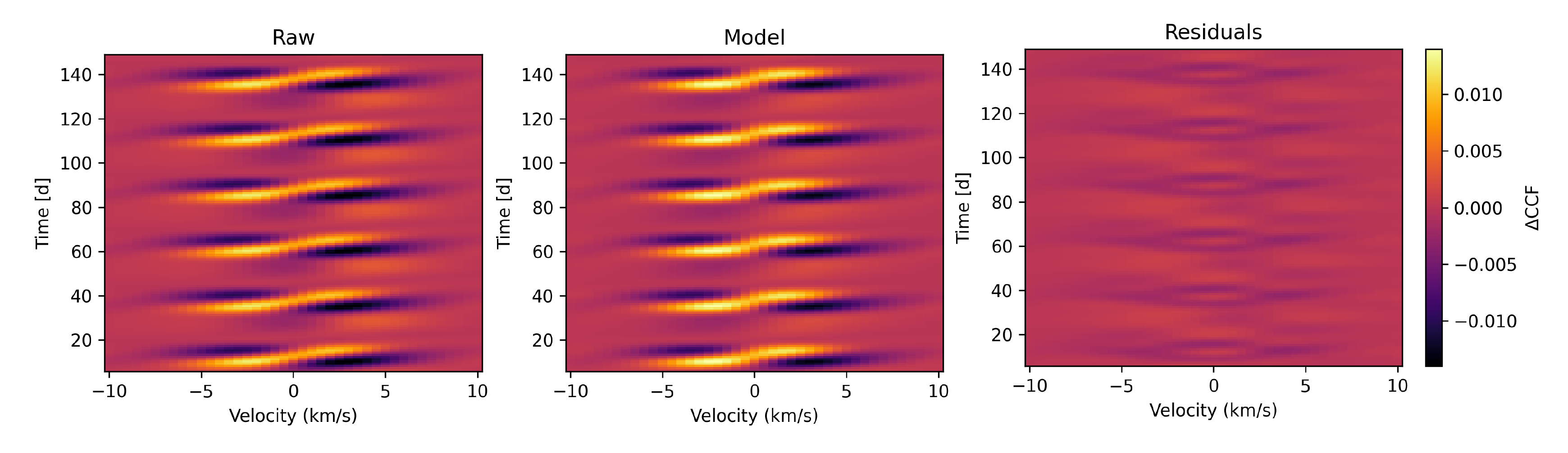}
    \caption{Left panel: simulated reference-subtracted CCF ($\Delta \rm{CCF}$) with \texttt{SOAP2} (case <I>). Middle panel: the best-fit CCF-GP model. Right panel: the residuals after subtracting the best-fit model.}
    \label{fig:ccf-test11}
\end{figure*}

\begin{figure*}
	\centering
	\includegraphics[width=1.0\textwidth]{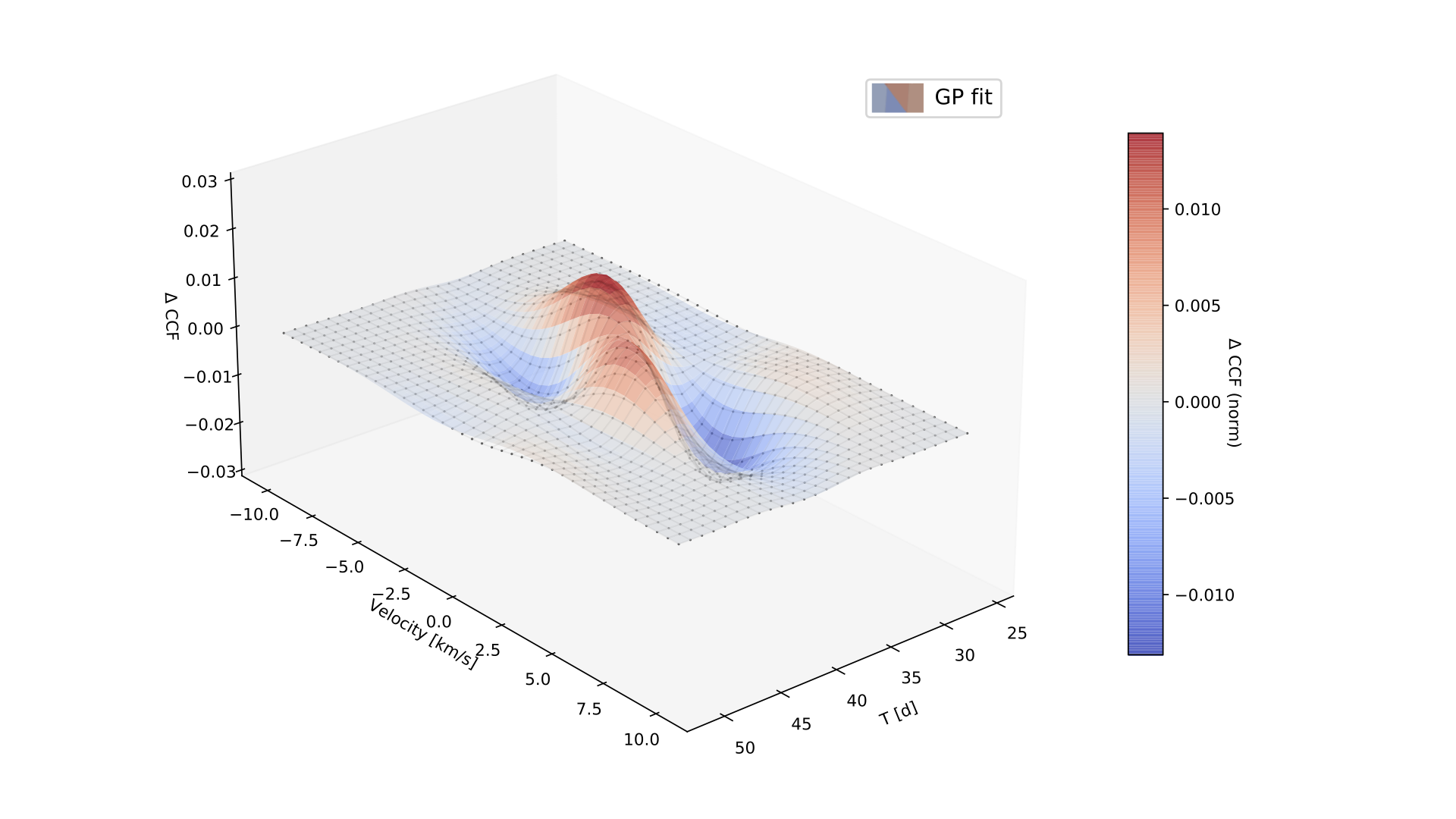}
    \caption{3-D visualization of the CCF-GP model fit to the referenced-subtracted CCF, $\Delta \rm{CCF}$ simulated with \texttt{SOAP2} (case <i>), zooming into one rotation period. The black markers represent the simulated data, and the coloured surface illustrates the model fit, with the colour gradient indicating $\Delta \rm{CCF}$ values (positive in red and negative in blue).}
    \label{fig:ccf-test12}
\end{figure*}

For the first case, we simulated a stellar surface featuring one spot and one facula, primarily to validate the framework's concept, specifically, its ability to track the signatures in the CCFs influenced by both the spot and facula. The simulated reference-subtracted CCFs ($\Delta \rm{CCF}$) are shown in Figure \ref{fig:ccf-test11} in the left panel, with the best-fit CCF-GP model in the middle panel and residuals in the right panel. The typical amplitude of the variations in the $\Delta \rm{CCF}$ is around $1\%$. Figure \ref{fig:ccf-test12} shows a 3-D representation of $\Delta \rm{CCF}$ versus velocity and time, enabling a clearer visualization of the model fit. This figure highlights that the variations are correlated in both the velocity and time domains, and demonstrates that their correlations are effectively captured by our framework.

\begin{figure}
	\centering
	\includegraphics[width=\columnwidth]{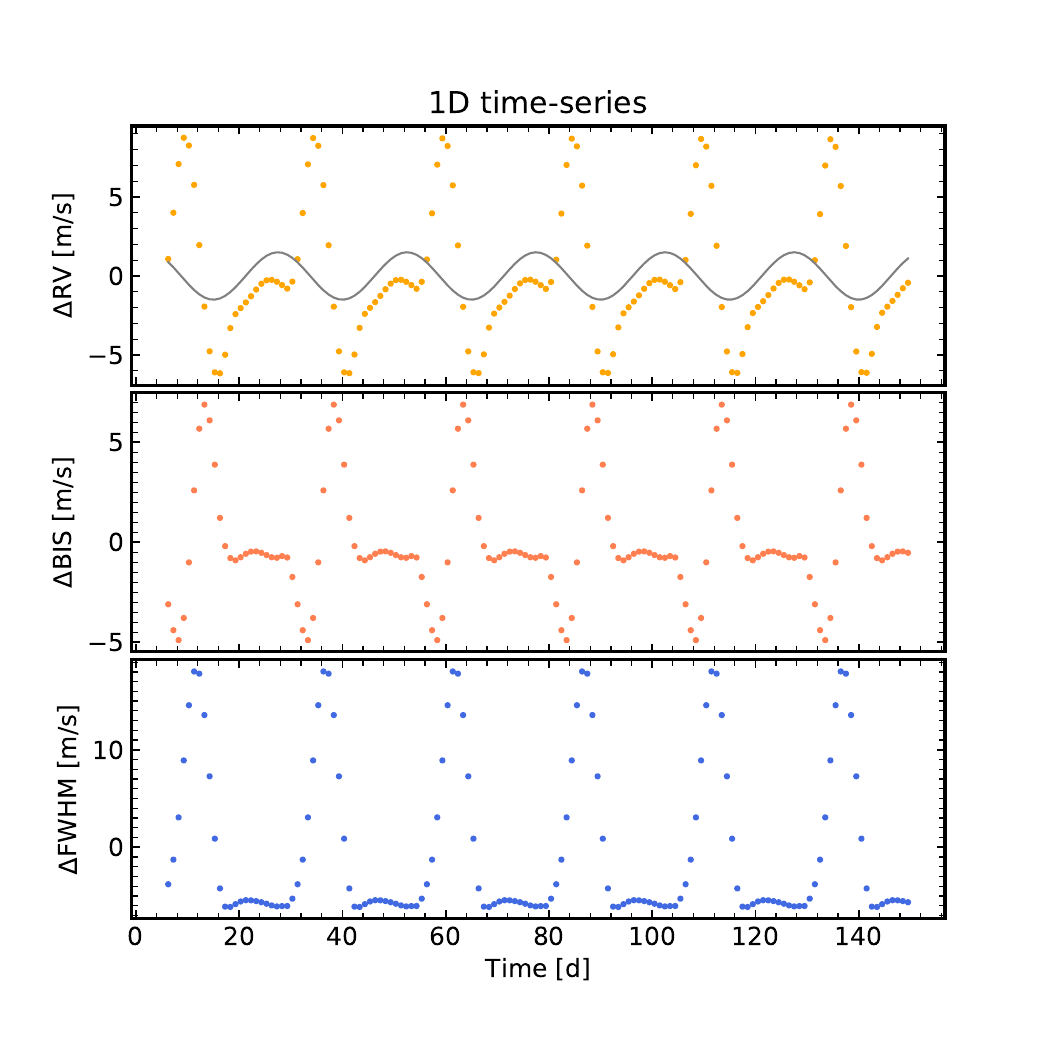}
    \caption{Mean-subtracted 1-D time-series, including RV, BIS, and FWHM, which are derived from the planet-injected simulated CCFs generated with \texttt{SOAP2} case <ii>. The first panel also shows the planet-induced RV in grey.}
    \label{fig:ccf-test21}
\end{figure}

For the second case, our objective was to demonstrate the model's ability to limit the degeneracy between the signals induced by stellar activity and planets. For this test, we simulated a more intricate and realistic stellar surface with two spots and two faculae, with the properties listed in Table \ref{tab:soap2}. To challenge the framework, we then injected a planetary signal into the simulated CCFs, with exactly the same orbital period as the star's rotation period (in this instance, 25 days). This planet signal had a semi-amplitude of 1.0~m/s and its phase was arbitrarily set to 0.3 $\rm{\pi \ rad}$. The phase is defined with reference to T=0, and phase=0 corresponding to the position of inferior conjunction for the injected planet. Figure \ref{fig:ccf-test21} displays the mean-subtracted 1-D proxies, including RV, BIS, and FWHM, derived from the planet-injected simulated CCFs. The first panel also shows the planet-induced RV in grey to enable its comparison to the overall RV signal.

\begin{figure}
	\centering
	\includegraphics[width=\columnwidth]{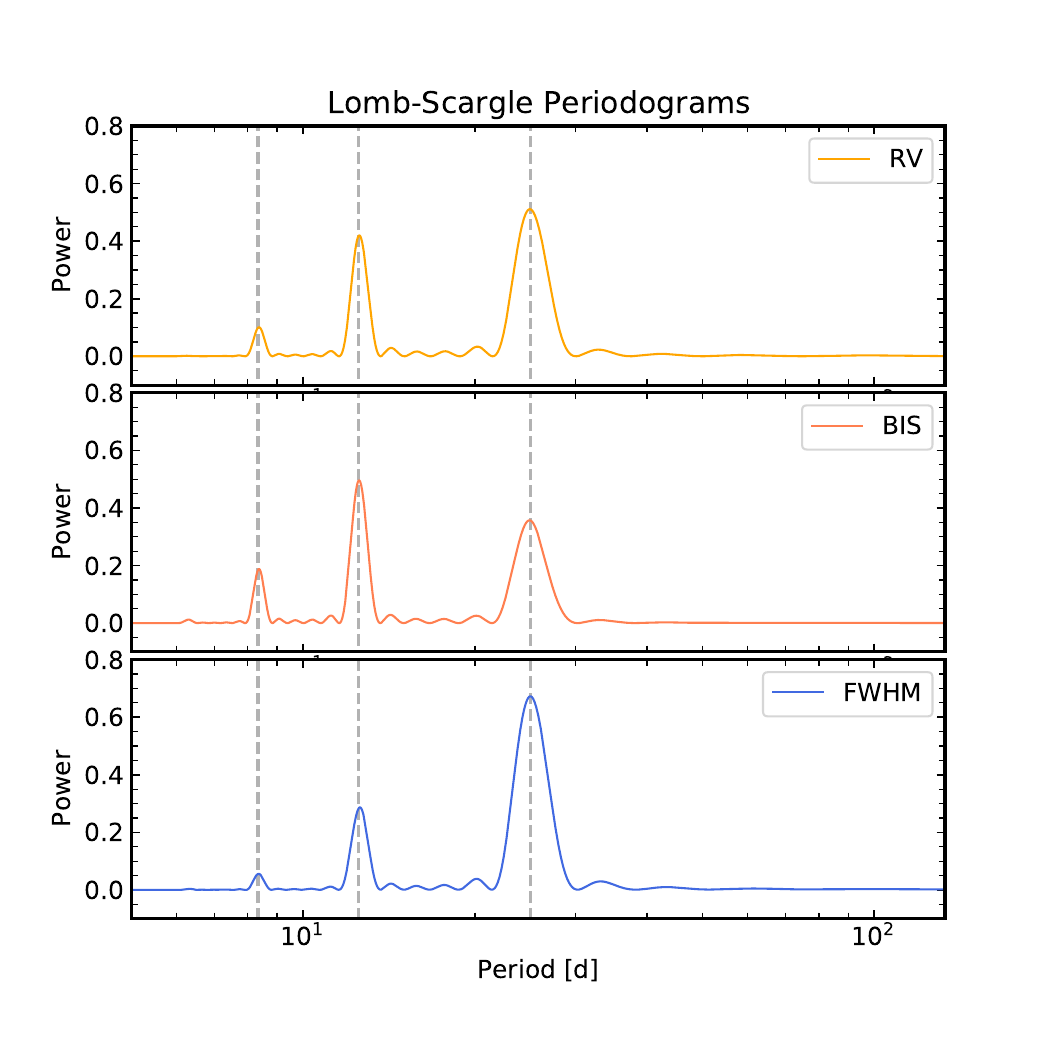}
    \caption{Lomb-Scargle periodogram of each 1-D time series (including RV, BIS, and FWHM) shown in Figure \ref{fig:ccf-test21}. The grey vertical markers indicate the positions of the 25-day rotation signal and its harmonics.}
    \label{fig:ccf-test22}
\end{figure}

We first computed the Lomb-Scargle periodogram (using \texttt{Astropy} \citep{astropy:2013, astropy:2018, astropy:2022} of each 1-D time series, as shown in Figure \ref{fig:ccf-test22}. The grey vertical markers indicate the positions of 25-day signals and their harmonics. Using the traditional approach of comparing the periodogram of the RV and 1-D activity indicators, any peak in the RV periodogram that also appears in the periodograms of the activity indicators would be dismissed as likely arising from activity. Therefore, discerning signals induced by planets from those induced by stellar activity is impossible using periodograms alone when both share the same period.

\begin{figure}
	\centering
	\includegraphics[width=0.8\columnwidth]{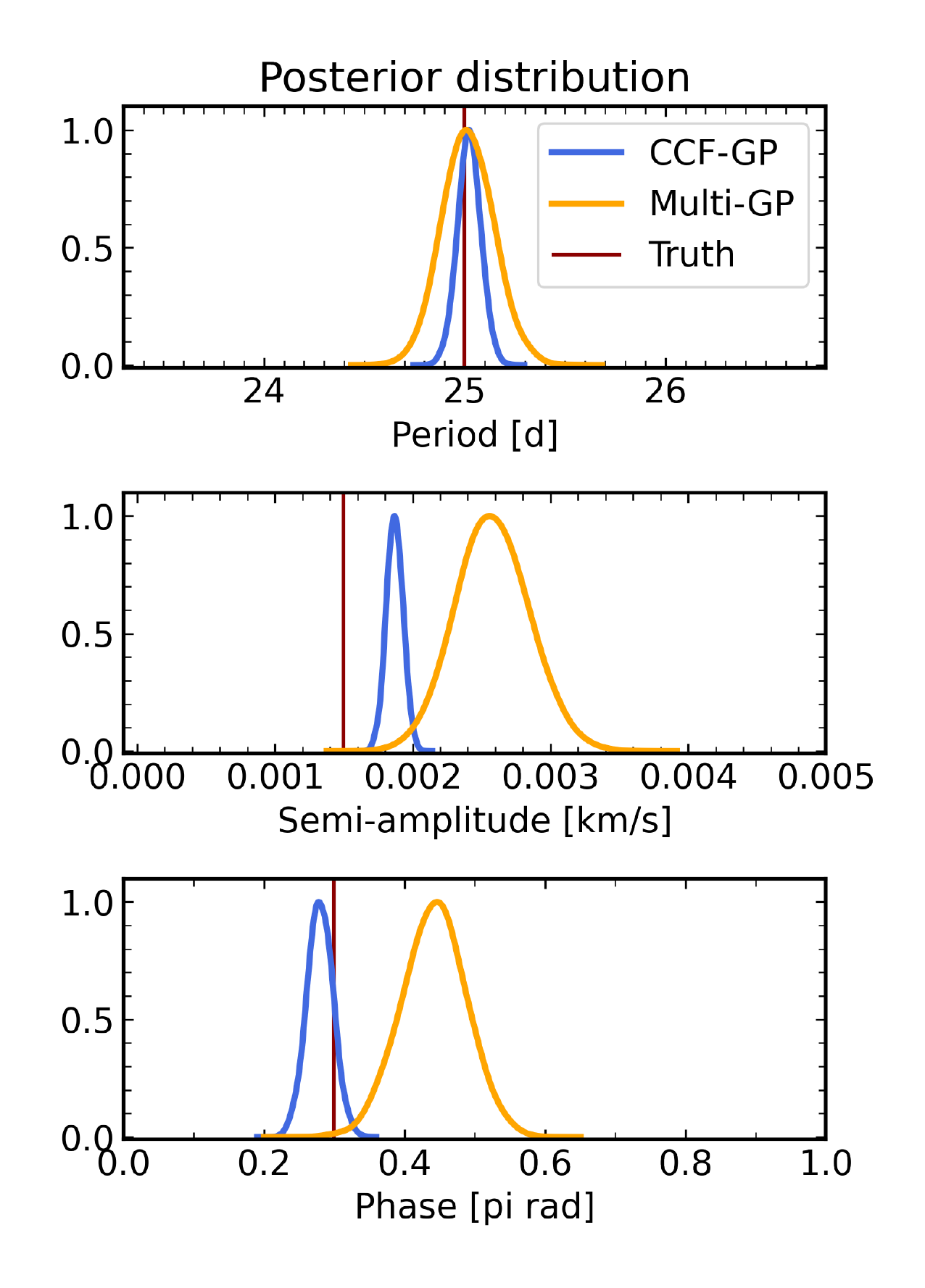}
    \caption{Posterior distributions of the recovered orbital parameters for the simulated dataset (\texttt{SOAP2} case <ii>) from both CCF-GP (in blue) and multi-GP (in orange) models. Vertical red lines indicate the true values from the injections.}
    \label{fig:ccf-test23}
\end{figure}

We subsequently applied the CCF-GP framework to the CCF time series. For comparison, we also applied the multi-GP framework, to the RV, BIS, and FWHM extracted from the same CCF time series. Both models incorporated the same planet model for RVs, featuring a single Keplerian. Crucially, the parameter priors in both the multi-GP and CCF-GP were identical and remained uninformative. Figure~\ref{fig:ccf-test23} presents the posterior distribution of the recovered orbital parameters from both CCF-GP (in blue) and multi-GP (in orange) models. Vertical red lines indicate the true values from the injections.

It is evident that both models adeptly recover the 25-day period. However, as the stellar rotation period is also at 25 days, to be certain that the detected signal is indeed that of the planet, we also required that the phase of the planet signal be retrieved. Observing the posterior distribution over the phase reveals a notable advantage for the CCF-GP, which captures the truth within a 1-$\sigma$ uncertainty range, while the phase determined via the multi-GP deviates by more than $0.1 \ \pi \ \rm{rad}$ from the true value. Regarding the semi-amplitude, both models exhibit a tendency to provide inflated estimates of $K=1.86_{-0.05}^{+0.05}$~m/s (CCF-GP) and $K=2.57_{-0.24}^{+0.26}$~m/s (multi-GP), though the CCF-GP derived value is closer to the truth of $K=1.5$~m/s. This inflation implies that the models are conservative in removing the activity signals, leaving a fragment of these signals intertwined with the planet signals. In the case of the CCF-GP model, this slightly conservative behaviour might come from the simplification of modelling the CCFs as vanilla Gaussian processes (as discussed earlier in Section \ref{sec:CCF-method}), potentially neglecting minor yet pertinent non-Gaussian aspects of the signals. 

With the same simulated dataset, we conduct evidence comparisons on combinations of model and simulated dataset with 0 or 1 planet with CCF-GP and multi-GP frameworks. The four cases are listed in Table \ref{tab:soap2_evi}, with their corresponding evidence $\log Z$ estimated from the nested sampling. We also provide $\Delta\log Z$, which are calculated as the evidence of the `correct' model minus the `incorrect' model. We find that for the data containing 1 injected planet, the model containing 1 planet (1-planet model) is always strongly preferred compared to the 0-planet model for both CCF-GP and multi-GP. However, for the data without planet injection, the 1-planet model is preferred to the 0-planet model for both CCF-GP and multi-GP, though such preference for the latter one is less presented. This suggests further avenues for improvement to prevent false-positive detections.

Overall, we find that in this specific test, the CCF-GP approach generally outperforms the multi-GP approach in disentangling activity and planetary signals, especially when their periods closely match the star's rotation period or its harmonics.

\begin{table*}
    \centering
    \caption{Evidence for combinations of model and simulated dataset with 0 or 1 planet, generated from \texttt{SOAP2} case <ii>.}
    \label{tab:soap2_evi}
    \begin{tabular}{ccccc}
       \hline
       Case  & CCF-GP $\log Z$ & CCF-GP $\Delta\log Z$ & Multi-GP $\log Z$ & Multi-GP $\Delta\log Z$\\
       \hline
       0-planet data and 0-planet model & $44232.20 \pm 0.33$ & $-45.44$ & $2348.28 \pm 0.30$ & $-2.31$ \\
       0-planet data and 1-planet model & $44277.64 \pm 0.32$ & -- & $2350.59 \pm 0.29$  & -- \\
       1-planet data and 0-planet model & $43941.19 \pm 0.24$ & -- & $2292.24 \pm 0.30$  & -- \\
       1-planet data and 1-planet model & $44242.19 \pm 0.34$ & $301.00$ & $2349.71 \pm 0.29$  & $57.47$ \\
       \hline
    \end{tabular}
\end{table*}

\subsection{Injection and recovery test on HARPS-N solar data} \label{sec:CCF-test-solar}
To assess the behaviour of the CCF-GP framework under more realistic conditions, and to determine its ability to detect low-amplitude planet signals (while having a benchmark truth), we conducted planet injection and recovery tests using HARPS-N solar data.

\begin{figure*}
	\centering
	\includegraphics[width=0.9\textwidth]{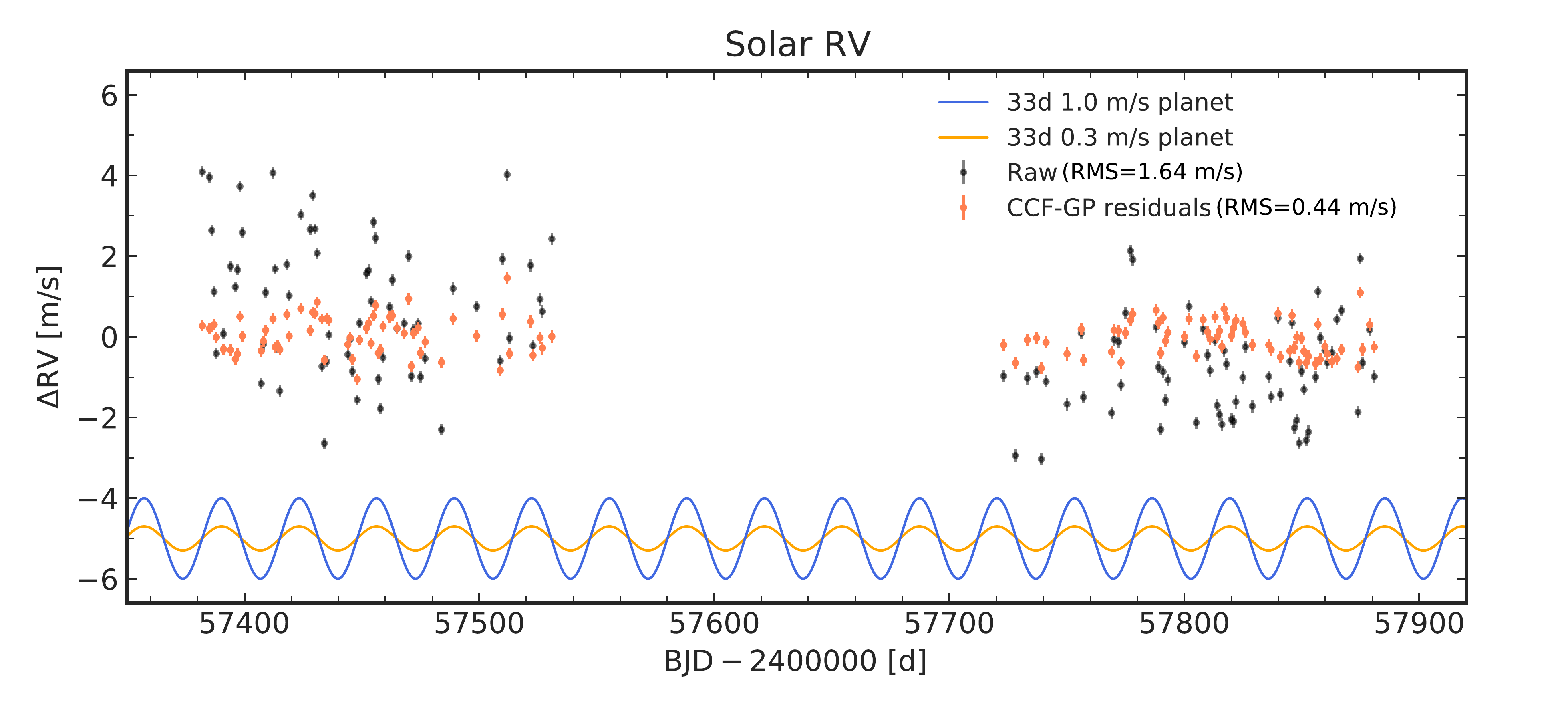}
    \caption{The HARPS-N solar RVs used for injection/recovery tests (black markers) and the residual RVs (coral markers), which were measured from the CCFs after subtracting the best-fit CCF-GP model. Also shown are the injected 33-d planet signals for both semi-amplitude settings in blue (1.0 m/s) and orange (0.3 m/s), offset by -5 m/s in the figure.}
    \label{fig:ccf-test31}
\end{figure*}

The spectra were collected using the HARPS-N solar telescope \citep{Cosentino2012,Dumusque2015,phillips2016} with observations taken for several hours each day at 5-minute cadence since 2015. Initial processing of the raw data was performed by \citet{Dumusque2021} employing the ESPRESSO data-reduction software \citep[DRS;][]{Pepe2021}. Following \citet{2019MNRAS.487.1082C}, cloud-affected spectra were excluded, and the remainder were shifted to the solar rest frame through interpolation using the Solar barycentric RV solution derived with the \texttt{JPL Horizons} software \citep{1996DPS....28.2504G}. All sets of three consecutive 5-minute exposures are combined to make one 15-minute exposure for mitigating the effect of p-mode oscillations. We then simulated the temporal sampling one might expect for hypothetical stars observed from La Palma by adding 0.5 days to the observation times (to shift them to night times), then selected at most one observation each night when a hypothetical star located at $\rm{RA}=90^{\circ}$, $\rm{DEC}=+30^{\circ}$ is observable at an airmass below 1.8. Our test utilized data from mid-2015 to mid-2017, corresponding to an active part of the solar magnetic activity cycle. There are 130 observations in total, with an average sampling rate of one point per 3.8 days, and a typical SNR of $\sim$370 per 0.82 km/s velocity bin in \'echelle order 60 of the spectra. Figure \ref{fig:ccf-test31} shows the mean-subtracted RV measured from the selected data, which exhibits a root mean square (RMS) value of 1.64~m/s.

We first applied the CCF-GP framework to analyse the HARPS-N solar CCFs without injecting any planetary signals. This involved fitting the CCF-GP model to the data and subsequently subtracting it to get the CCF residuals. We then computed the RV residuals by fitting a Gaussian function to each of these residual CCFs, which are represented as coral markers in Figure \ref{fig:ccf-test31}. The RV residuals have a RMS value of 0.44 m/s, with an RMS reduction factor of around 4 when compared to the raw solar RVs. 

\begin{figure*}
	\centering
	\includegraphics[width=0.9\textwidth]{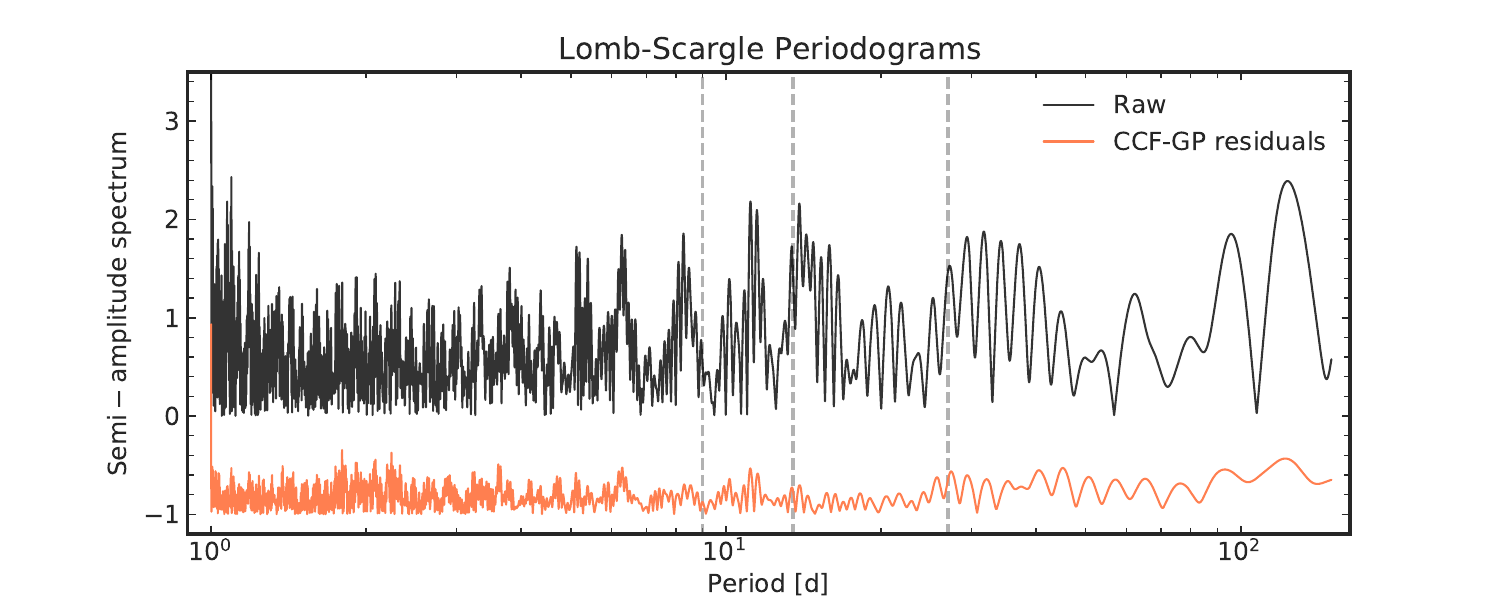}
    \caption{Lomb-Scargle periodogram of the raw solar RV and the RV residuals shown in Figure \ref{fig:ccf-test31}. The grey vertical markers indicate the positions of the 27-day solar-rotation signal and its harmonics.}
    \label{fig:ccf-test31p}
\end{figure*}

We also calculated the Lomb-Scargle periodograms of the raw and the residual RVs, shown in Figure \ref{fig:ccf-test31p}. The periodograms are normalized as half of the square root of the power spectrum density (PSD) values, and are thus called the semi-amplitude spectrum. Grey vertical markers indicate the positions of the 27-day solar-rotation signal and their harmonics. Most of the frequency components, especially the ones related to the solar rotation period, are significantly reduced in the residual RVs.

We tested two scenarios by injecting a planet into the HARPS-N solar CCFs with semi-amplitudes of 1.0 m/s and 0.3 m/s. In both cases, the orbital period was set to 33 days, and the phases were arbitrarily set to $0.7 \ \pi \ \rm{rad}$. The phase is defined with reference to BJD=2450000. These correspond to planets of masses around 5 and 1.5 Earth masses respectively. Figure \ref{fig:ccf-test31} shows the RVs of the planet for both semi-amplitude settings in blue and orange. We then applied the CCF-GP framework on the planet-injected CCF. All priors on the free parameters are uninformative, which means this is a blind search (though we only tested a one-planet model). 

\begin{figure*}
	\centering
	\includegraphics[width=0.6\textwidth]{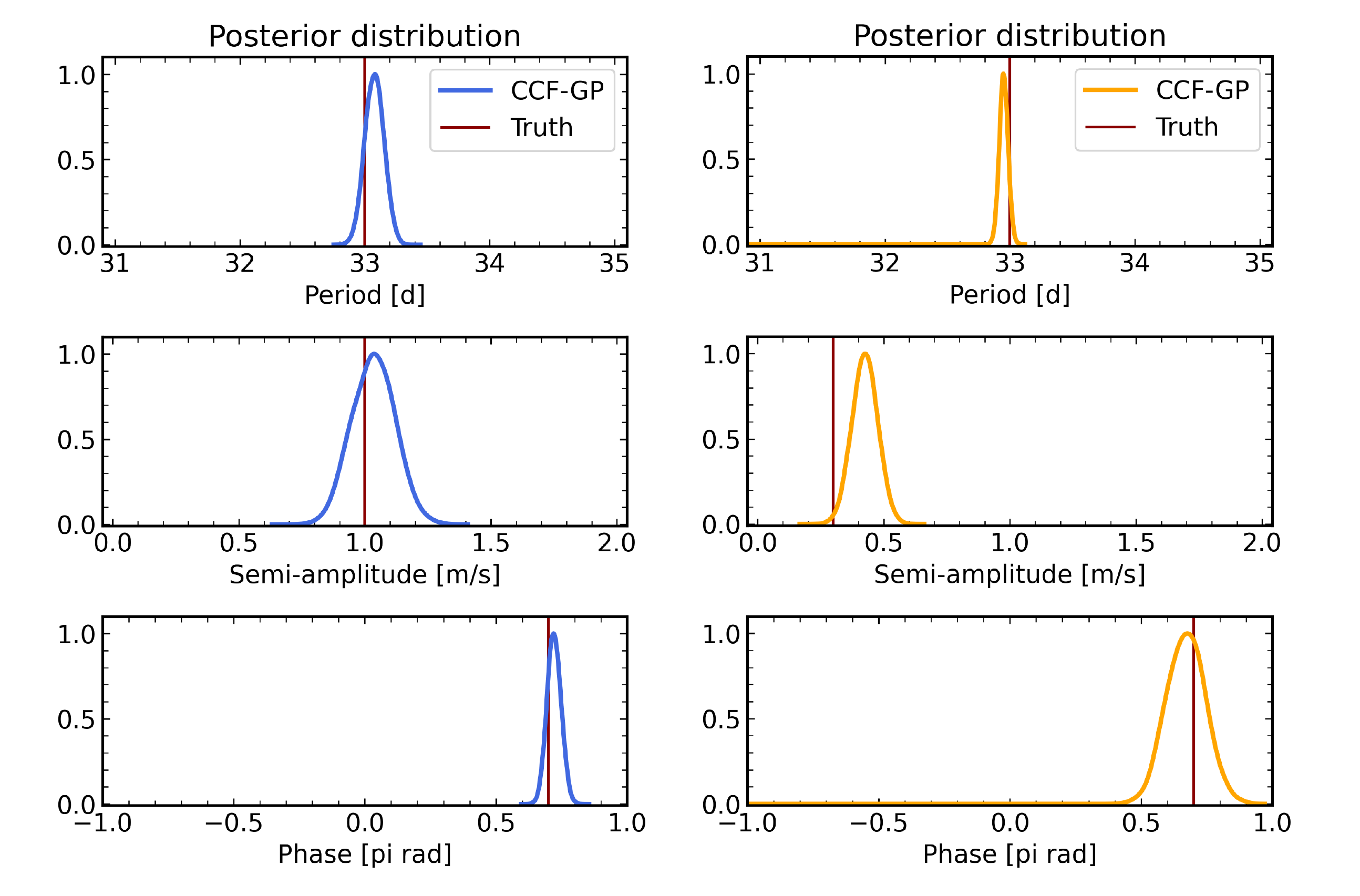}
    \caption{Posterior distribution of the extracted planetary parameters for the HARPS-N injection/recovery tests for planets with 33-d period and semi-amplitudes of 1.0 m/s (left, blue) or 0.3 m/s (right, orange). Vertical red lines indicate the true values from the injections.}
    \label{fig:ccf-test32}
\end{figure*}

Figure \ref{fig:ccf-test32} shows the posterior distribution of the extracted planetary parameters, with vertical red lines indicating the true values from the injections. For the 1.0 m/s case, all the planet parameters are accurately recovered. For the 0.3 m/s case, while the period and phase are well-retrieved, the semi-amplitude starts to be overestimated, with a derived value of $K=0.42_{-0.04}^{+0.04}$~m/s. This discrepancy can be attributed to multiple factors. As already discussed in Section \ref{sec:CCF-test-soap}, one plausible explanation is the overlooked non-Gaussianity of the CCF time series. Another factor is the presence of physical effects in the solar data that are not accounted for in our model, such as the bisector shape alterations at active regions, and the effects of super-granulation, etc.

We also tested a case specifically targeting long-period planets. For this case, we utilized the same HARPS solar RV dataset as in the previous cases, but without accounting for the hypothetical star's observability constraints. The data contains 290 observations, with an average sampling rate of one data point per 1.7 days. This number of observations would take around four or five years to gather on the hypothetical star considered in the previous test. Once a full decade of HARPS-N solar data becomes publicly available, we will be able to simulate observations from a survey such as the Terra Hunting Experiment \citep{2018MNRAS.479.2968H} more realistically.

\begin{figure*}
	\centering
	\includegraphics[width=1.0\textwidth]{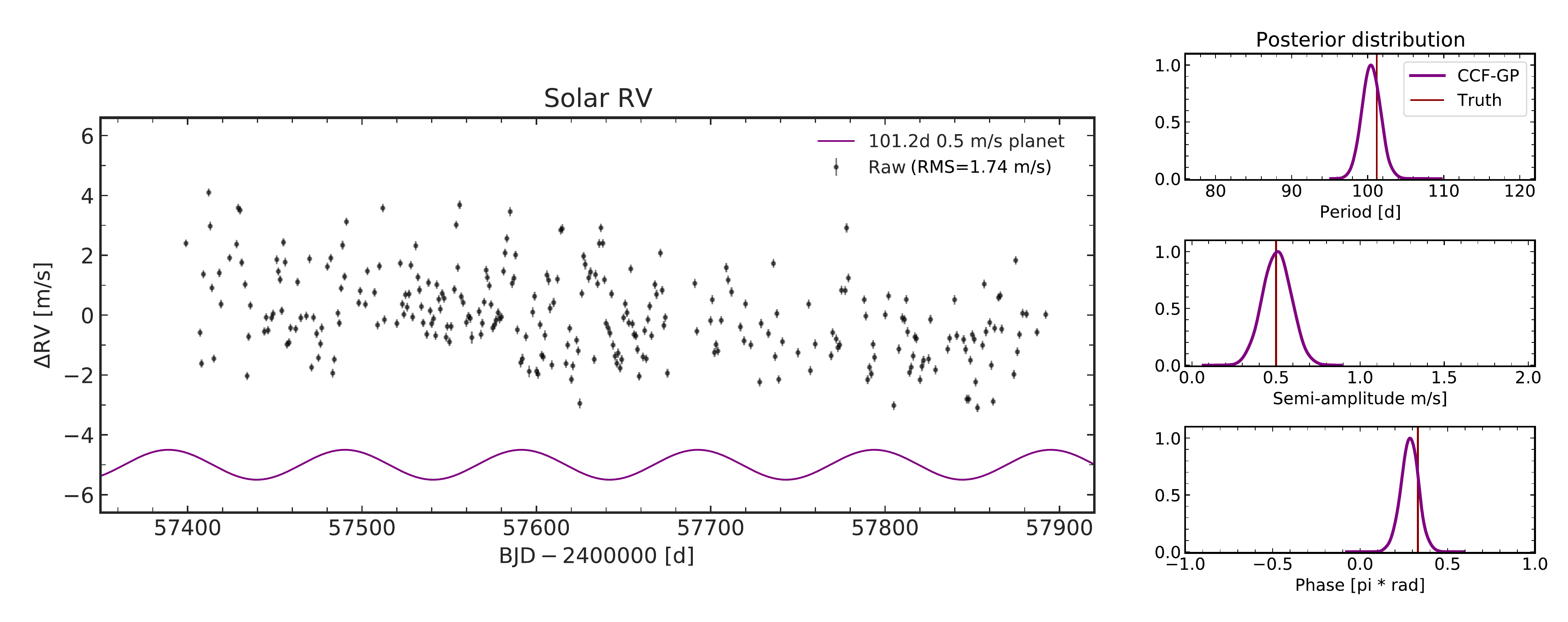}
    \caption{As Fig. \ref{fig:ccf-test31} and \ref{fig:ccf-test32}, but the figure shows the injection/recovery test of the long-period planet at 101.2-d orbital period, and of a semi-amplitude of 0.5 m/s.}
    \label{fig:ccf-test311}
\end{figure*}

For this test, we injected a planet with an orbital period of 101.2 days and a semi-amplitude of 0.5 m/s, corresponding to a planet of 3.6 Earth masses. The corresponding raw and planet RVs are shown in the left panel of Figure \ref{fig:ccf-test311}. The results of applying the CCF-GP model, as shown in the right panel of the same figure, reveal that the recovered planet parameters align well with the true values.

\subsection{Application to real observations} \label{sec:CCF-test-real}

\begin{figure}
	\centering
	\includegraphics[width=\columnwidth]{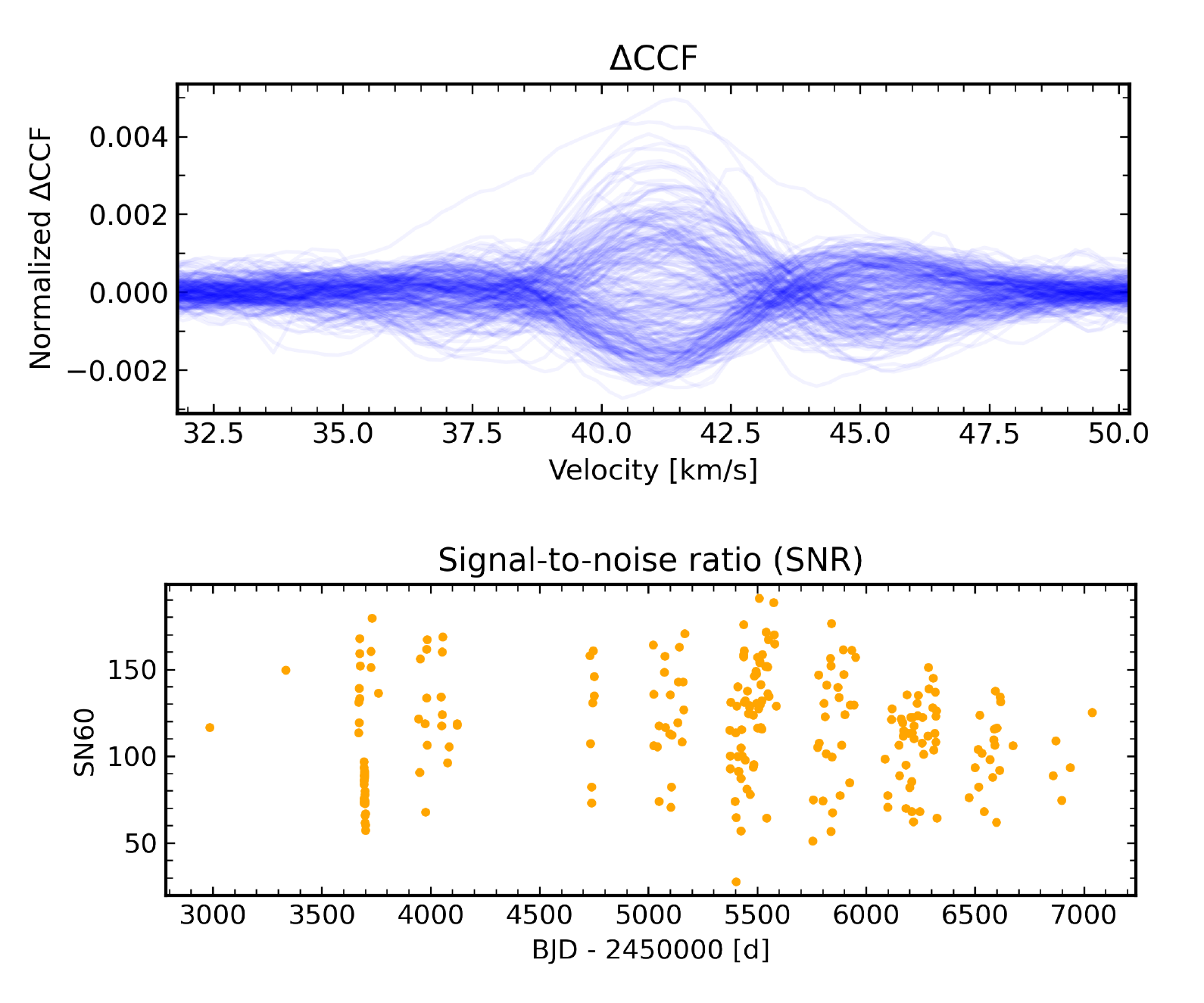}
    \caption{Reference-subtracted CCFs and the associated SNR per pixel (in échelle order 60) plotted over time. The SNR for these observations varies between approximately 50 and 180, with a median value of 100.}
    \label{fig:ccf-test41}
\end{figure} 

\begin{figure*}
	\centering
	\includegraphics[width=0.7\textwidth]{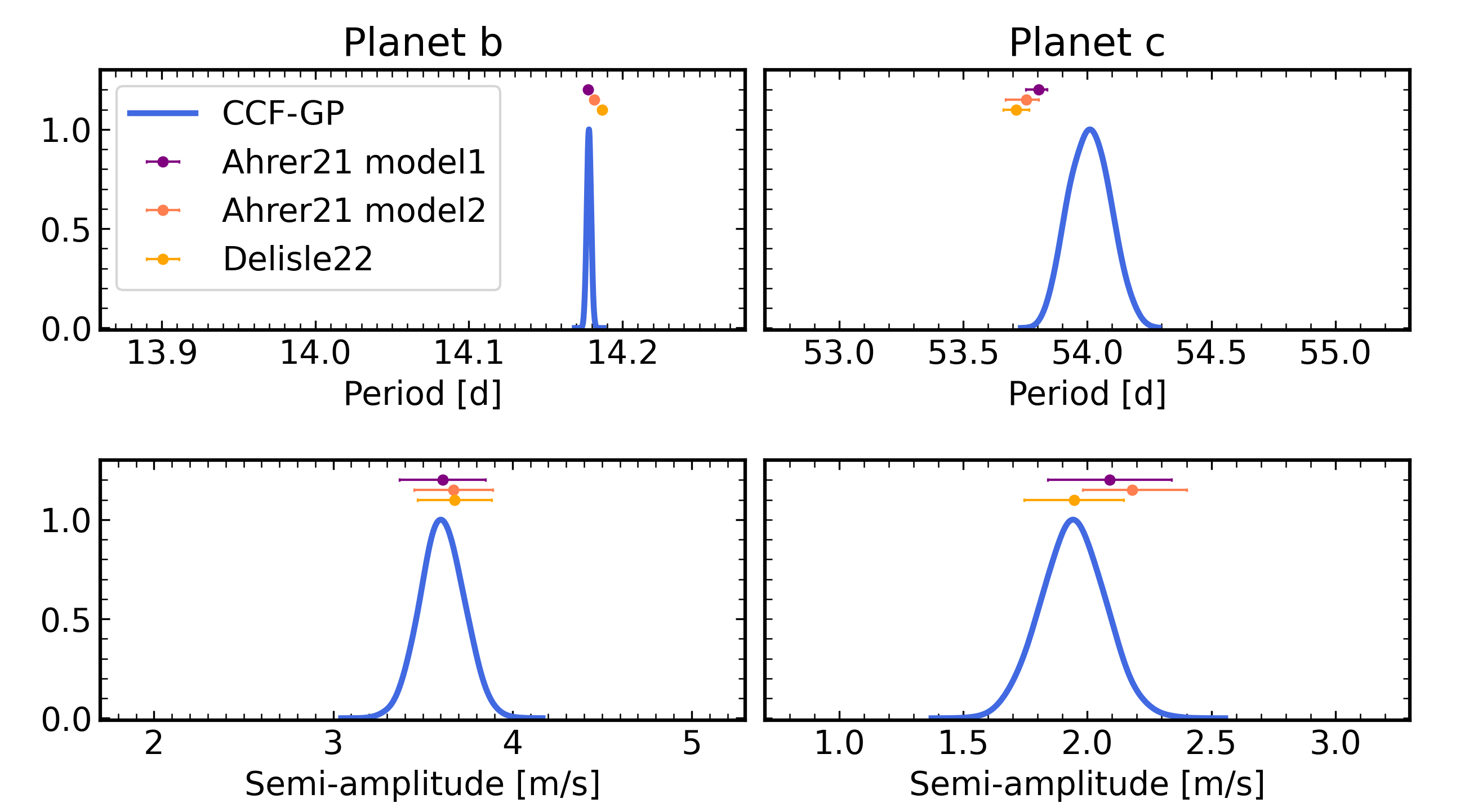}
    \caption{Posterior distributions of the derived planetary parameters from the CCF-GP analysis on the HD~13808 data. The derived values and associated uncertainties of the corresponding parameters from \citet{Ahrer2021} and \citet{D22} are indicated by the markers with errorbars.}
    \label{fig:ccf-test42}
\end{figure*} 

We applied the CCF-GP framework to HARPS observations of HD~13808 to demonstrate its efficacy on real-world observations with an average SNR of $\sim$100. HD~13808 is a system consisting of a K2V dwarf star orbited by two confirmed non-transiting Neptune-mass planets \citep{Ahrer2021}. The data, which are queried from the ESO archive, include 246 spectroscopic measurements taken with HARPS, mostly between 2005 and 2016. The data underwent pre-processing via the automated DRS pipeline. Figure \ref{fig:ccf-test41} presents the series of reference-subtracted CCFs and the associated SNR per pixel (in echelle order 60). The SNR for these observations varies between approximately 50 and 180, with a median value of around 100. We applied the CCF-GP framework, incorporating a two-planet RV model. As in previous tests, all priors on free parameters remained uninformative, implying that this was a blind detection test.

In Figure \ref{fig:ccf-test42}, the posterior distributions of the extracted planetary parameters are displayed. For real observations, a definitive `true' value remains elusive. Therefore, we compared the extracted values with those reported in two recent publications, \citet{Ahrer2021} and \citet{D22}. \citet{D22} implemented a modified version of the multi-GP with a fast MEP kernel. Their best-fit values for the planet parameters are indicated by the orange markers. Conversely, \citet{Ahrer2021} evaluated various models for activity, endorsing that the harmonic activity model (model 1 in Figure \ref{fig:ccf-test42}), which simultaneously fits the 1st, 2nd, and 3rd harmonic of rotation period to the BIS and RVs, and the multi-GP model (model 2) as two of the most robust models; their results are indicated by purple and coral markers. The derived semi-amplitudes for planets b and c are $K_{\rm b}=3.60_{-0.11}^{+0.12}$~m/s and $K_{\rm c}=1.94_{-0.12}^{+0.12}$~m/s, which are within one sigma deviation from the values reported in \citet{Ahrer2021} and \citet{D22}. This example demonstrates that the CCF-GP framework can effectively work with data of averaged SNR of around 100, accommodating the majority of current observations and producing results consistent with existing approaches.


\section{Conclusions} \label{sec:conc}
In this paper, we introduced the CCF-GP framework, a novel approach developed to model stellar activity signals within time-series of CCFs using GPs. This model addresses the distortions in CCFs due to activity regions on the stellar surface, i.e., spots and faculae, stemming from both the photometric effect and the inhibition of convective blueshift effect.

Upon testing on both synthetic and real datasets -- with SNRs down to around 100 -- the framework showcased its robustness in distinguishing between activity and planetary signals, even when they share identical periods. Planet injection/recovery tests on the realistically sampled HARPS-N solar data highlight the framework's potential to push the detection limit of planetary signals down to $\rm{K=0.3~m/s}$ at $\rm{P=33~d}$ (corresponding to a planet of 1.5-Earth mass) during the Sun's high activity phase in its magnetic cycle. The framework surpasses the performance of multi-GP models on RV and activity indicators in the tests conducted. We note that similar to the multi-GP approach, CCF-GP also requires adequate temporal sampling relative to the rotation period of the star for optimal performance.

In the future development of this model, we plan to explicitly address the issue of non-Gaussianity mentioned in Section \ref{sec:CCF-GP}. One approach as a substitute for using the current approximate GP model is to place a GP prior on $F$ and use, e.g., Markov chain Monte Carlo (MCMC) methods to sample from it and then propagate the samples to approximate the exact stochastic process of the CCF model. However, this comes at the cost of increased computational demands. 

Additionally, there is significant potential in enriching the model with more detailed physics. One would be the replacement of the current QP or MEP kernels with kernels that are obtained from physical parameters, such as the `impulse response' GP kernel developed in \citet{Hara2023}. Furthermore, the model could benefit from incorporating the differential responses of spectral lines to activity, as well as the effects of super-granulation. The inherent flexibility of the model's mathematical framework ensures that these integrations can be implemented seamlessly.

\section*{Acknowledgements}
H.Y., S.A., B.K. acknowledge funding from the European Research Council under the European Union’s Horizon 2020 research and innovation programme (grant agreement No 865624, GPRV). M.C. acknowledges the SNSF support under grant P500PT\_211024.

\section*{Data Availability}
This work was based entirely on publicly available data downloaded from the Data Analysis Center for Exoplanets (DACE)\footnote{\url{https://dace.unige.ch/dashboard/}} and the ESO Science Archive\footnote{\url{https://archive.eso.org/cms.html}}.

\bibliographystyle{mnras}
\bibliography{ref} 

\bsp	
\label{lastpage}
\end{document}